\documentclass[aps,prb,twocolumn,floatfix]{revtex4-2}
\usepackage{graphicx}
\usepackage{mathtools}
\usepackage[normalem]{ulem}
\usepackage{dsfont}
\usepackage[colorlinks=true,linkcolor=black,urlcolor=black,citecolor=black]{hyperref}
\usepackage{bm}

\usepackage{changes}

\usepackage{microtype}

\newcommand*{\spinop}[1]{\bm{{#1}}} % Bold operator with hat

\DeclareMathOperator{\Tr}{\rm Tr}
\newcommand{\blangle}{\bigl\langle}
\newcommand{\brangle}{\bigr\rangle}
\newcommand{\dlangle}{\langle\kern-1.5pt\langle}
\newcommand{\drangle}{\rangle\kern-1.5pt\rangle}
\newcommand{\bdlangle}{\blangle\kern-3pt\blangle}
\newcommand{\bdrangle}{\brangle\kern-3pt\brangle}
\newcommand*{\braa}[1]{\langle{#1}|}
\newcommand*{\kett}[1]{|{#1}\rangle}
\newcommand*{\average}[1]{\langle{#1}\rangle}

\newcommand*{\expect}[3]{\langle{#1}|{#2}|{#3}\rangle}

\begin{document}
\title{Stability of a quantum skyrmion: projective measurements and the quantum Zeno effect}

\author{Fabio Salvati}%
  \email{fabio.salvati@ru.nl}%
\author{Mikhail I. Katsnelson}%
\author{Andrey A. Bagrov}%
\author{Tom Westerhout}%
  \affiliation{Institute for Molecules and Materials, Radboud University, Heijendaalseweg 135, 6525 AJ Nijmegen, The Netherlands}

\author{}

\begin{abstract}
Magnetic skyrmions are vortex-like quasiparticles characterized by long lifetime and remarkable topological properties.
That makes them a promising candidate for the role of information carriers in magnetic information storage and processing devices.
Although considerable progress has been made in studying skyrmions in classical systems, little is known about the quantum case: quantum skyrmions cannot be directly observed by probing the local magnetization of the system, and the notion of topological protection is elusive in the quantum realm.
Here, we explore the potential robustness of quantum skyrmions in comparison to their classical counterparts.
We theoretically analyze the dynamics of a quantum skyrmion subject to local projective measurements and demonstrate that the properties of the skyrmionic quantum state change very little upon external perturbations.
We further show that by performing repetitive measurements on a quantum skyrmion, it can be completely stabilized through an analogue of the quantum Zeno effect.

\end{abstract}
\maketitle

\section{Introduction}

Magnetic skyrmions are nanometer-size topological spin textures with integer topological charges and long lifetime~\cite{nagaosa_skyrm,takashima_skyrm, ochoa_skyrm,muhlbauer_skyrm, neubauer_skrym, romming_skyrm}.
Topological Hall effect, and current-driven motion of skyrmions with low-power consumption can be exploited for future applications of memory devices as well as information carriers~\cite{jonietz_spintransfertorque,fert_skyrm}.
The general contemporary tendency to ultra-miniaturization of such elements results naturally in attempts to study the regime when quantum effects become decisive, and that brings us to the topic of quantum skyrmions~\cite{lohani_skyrm,sotnikov_skyrm,garanin_skyrm,lorente_skyrm,siegl_skyrm,maeland_skyrm,mazur_jpsj,ernst_skyrm,sotnikov_darwinism}.
Contrary to classical skyrmions, quantum skyrmions are not topologically protected in a rigorous sense by the existence of a conserving topological charge.
Nonetheless, a quantum analogue of the topological invariant has been proposed~\cite{sotnikov_skyrm, sotnikov_darwinism}.
By studying the scalar chirality, defined as a local three-spin correlation function, it is indeed possible to characterize a quantum skyrmion, but the scalar chirality is, formally, not a topological charge~\cite{mermin}.
The relation between a quantum skyrmion state and its classical counterpart turns out to be highly nontrivial~\cite{sotnikov_darwinism}, and the robustness of the quantum skyrmion phase is not guaranteed.
However, for any potential applications, the robustness property is crucial, and its analysis in the quantum case deserves special attention.

In this paper, we study the effect of a few consecutive projective measurements on the stability of a quantum skyrmion state on the 19-site triangular lattice~\cite{sotnikov_skyrm,ernst_skyrm,sotnikov_darwinism}.
We demonstrate the robustness of the quantum skyrmion in terms of chirality as well as the spin-spin correlation function, which is, in principle, easier accessible experimentally.
This is an encouraging result since a projective measurement~\cite{vonneumann} is an idealization of the real physical process of reading information from a quantum device.
Our results therefore suggest that there is potential in using quantum skyrmions for storing information.

Another aim of our research is to use quantum skyrmions as a model system to study the general issues of the theory of quantum measurements.
Here, we focus on one concept from this theory, namely, the quantum Zero effect~\cite{sudarshan_zeno,facchi_zeno,aftab_zeno}, which is a counterintuitive effect of repeating projective measurements stabilizing an excited state of the quantum system.
Discreteness of the energy spectrum of the system is very important for the quantum Zeno effect, and its applicability for a 19-site quantum skyrmion, that is large enough to consider its spectrum quasi-continuous, is not clear.
We show that, nevertheless, an analogue of the quantum Zeno effect exists in this situation, and discuss possible reasons.

\section{Dynamics of a quantum skyrmion}

\subsection{Model}

In this work, we consider the quantum spin Heisenberg model with Dzyaloshinskii-Moriya interaction (DMI) and external magnetic field.
Its Hamiltonian reads:

\begin{equation}
    \label{eq:hamiltonian}
    {H} = \sum_{\left \langle i,j \right \rangle} J_{ij}\, \spinop{S}_i \cdot \spinop{S}_j + \sum_{\left \langle  i,j  \right \rangle} \mathbf{D}_{ij} \cdot \left[ \spinop{S}_i \times  \spinop{S}_j \right] + \sum_i B\, {S}^z_i,
\end{equation}

where $\spinop{S}_i = \frac{1}{2} \left(\sigma_i^x,\, \sigma_i^y,\, \sigma_i^z\right)$, $\sigma_i^\alpha$ for $\alpha \in \{x, y, z\}$ are the Pauli matrices on the $i$'th site, and $\langle i, j \rangle$ denote the nearest neighbors.
$\mathbf{D}_{ij}$ is an in-plane vector perpendicular to the bond $(i, j)$.
We will focus on the case of zero temperature, that is, we will work with the ground state of the Hamiltonian.
We will characterize the quantum skyrmion by the quantum scalar chirality (further called just chirality, for brevity):

\begin{equation}
    \label{eq:chirality}
    Q = \langle \hat{Q} \rangle = \frac{1}{\pi} \sum_{\langle ijk \rangle } \blangle   \spinop{S}_i \cdot [   \spinop{S}_j \times   \spinop{S}_k ] \brangle \,,
\end{equation}

where the sum runs over all non-overlapping triangular plaquettes.
The quantity $Q$ is a local three-spin correlation function defined on the neighboring lattice sites, and it was introduced~\cite{sotnikov_skyrm, berg_chirality} as the quantum analogue of the skyrmion topological index:

\begin{equation*}
    Q_\mathrm{top} = \frac{1}{4\pi} \int \bm{\mathrm{m}} \cdot \left[ \partial_x \bm{\mathrm{m}} \times \partial_y \bm{\mathrm{m}} \right] \mathrm{d}x\, \mathrm{d}y.
\end{equation*}

\begin{figure}
    \centering
    \includegraphics[width=\linewidth]{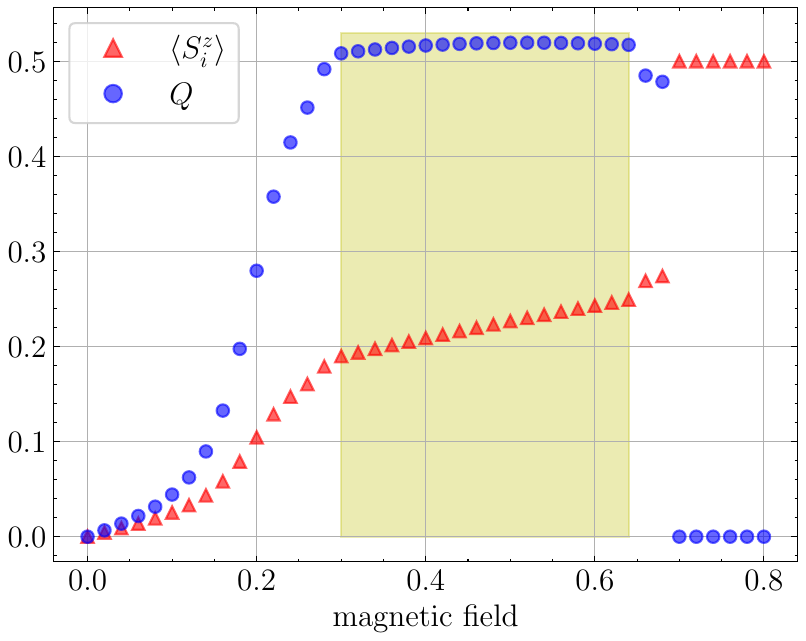}
    \caption{%
        Ground state phase diagram of the Hamiltonian~\eqref{eq:hamiltonian} on the 19-site triangular lattice with periodic boundary conditions.
        The chirality $Q$ is shown in blue circles, and the magnetization $\average{S_i^z}$ is shown in red triangles, both as functions of the external magnetic field $B$.
        In the ground state, $\average{S_i^z}$ is the same for every site $i$.
        The quantum skyrmion phase is highlighted in yellow.
    }
    \label{fig:topology}
\end{figure}

However, $Q$ is not a topological invariant in a mathematically rigorous sense.
Physically speaking, it is subjected to quantum spin fluctuations.
Nevertheless, it does characterize the quantum skyrmion phase since in the corresponding state, it displays unambiguously a non-zero and nearly constant value as a function of the external parameters, see Fig.~\ref{fig:topology}.

Exact diagonalization code~\cite{tom_ls} has been applied to solve numerically the Hamiltonian~\eqref{eq:hamiltonian} for a 19-site triangular lattice with periodic boundary conditions.
Following the previous works~\cite{sotnikov_skyrm,ernst_skyrm}, we set $J_{ij} = J = -0.5 D$, where $D = 1$ is the length of the DMI vectors $\mathbf{D}_{ij}$. Fig.~\ref{fig:topology} is in excellent agreement with the previous studies, but we nevertheless show it here as a basis for further calculations.

In Fig.~\ref{fig:topology}, we show the dependence of the chirality on the external magnetic field. For $0.30 \leq B \leq 0.64$, $Q$ is non-zero and nearly constant, and we associate this region with the quantum skyrmion phase.
Below $B = 0.30$, the system will slowly approach the helical spin state configuration at $B = 0$, where both chirality and magnetization go to zero.
Above $B = 0.64$, a first-order transition happens, and the system becomes a saturated ferromagnet.
Full characterization of these phases can be found in Ref.~\cite{sotnikov_skyrm}, and in this work, we use these results as a starting point for studying the robustness of a quantum skyrmion state.

\subsection{Quench dynamics following a projective measurement}

\begin{figure}
    \centering
    \includegraphics[width=0.7\linewidth]{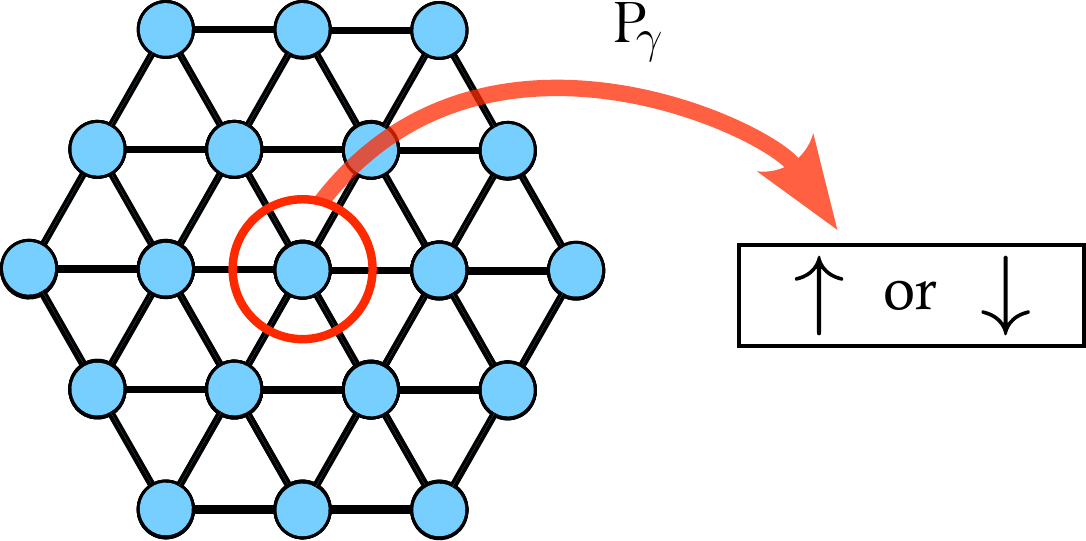}
    \caption{%
        Illustration of the projective measurement.
        The projection operator $P_\gamma$ (where $\gamma$ is either $\uparrow$ or $\downarrow$) acts on the center site of the lattice, and the center spin becomes either $|\!\uparrow\rangle$ or $|\!\downarrow\rangle$ after the projection.
    }
    \label{fig:lattice_projection}
\end{figure}

In any practical sense, robustness of a physical system should be understood as (partial) stability of its relevant properties under external deformations and environmental effects.
For a quantum skyrmion, although there is no notion of topological protection, if it is possible to show that its characteristic features do not change significantly upon strong perturbations, one can claim that the quantum skyrmion indeed shares the property of stability with its classical counterpart.
In light of great hope that skyrmions can provide a physical ground for dense information storage, it is natural to analyze their stability upon the act of reading information.
The latter, in its minimal form, can be modelled as a projective measurement of the $\sigma_z$-component of a single spin~\cite{chowdhury_projmeas, jordan_weakmeas, lloyd_weakmeas}, Fig.~\ref{fig:lattice_projection}.

Formally, the single-site projective measurement~\cite{vonneumann, sakurai, cirac_projection, gleason_projection, hylke_decoherence, hylke_pointer} can be described with the following operator acting on the quantum skyrmion ground state wave function $|\psi_{GS} \rangle$:
\begin{equation*}
    P^k_\gamma = \mathds{1}_2 \otimes \dots \otimes {\kett{\gamma}\braa{\gamma}}_k \otimes \dots \otimes \mathds{1}_2,
\end{equation*}
where $k$ is the index of the measured site, $\gamma$ is either $\uparrow$ or $\downarrow$, and $\mathds{1}_2$ is the $2\times 2$ identity matrix.

Immediately after the projective measurement, the $k$-th spin of the cluster will be oriented along the $z$ axis, as shown in Fig.~\ref{fig:lattice_projection}.
In what follows, we will always consider the measurement of the central site and will omit the index $k$ in the expressions.

Since $\kett{\psi_{GS}}$ is a pure state, its density matrix is simply $\rho = \kett{\psi_{\text{GS}}}\braa{\psi_{\text{GS}}}$.
Then, by denoting state right after the measurement as
\begin{equation*}
\kett{\psi_\gamma} \equiv \frac{P_\gamma \kett{\psi_{GS}}}{ \sqrt{\expect{\psi_{GS}}{P_\gamma}{\psi_{GS}}} }\;,
\end{equation*}
and according to the von Neumann theory of measurements~\cite{vonneumann}, the two possible outcomes of the measurement can be combined into the new density matrix:
\begin{equation}
    \label{eq:density_operator}
    \rho = \sum_{\gamma \in \{\uparrow, \downarrow\}} p_\gamma
    \kett{\psi_\gamma} \braa{\psi_\gamma}\,,\;\; \text{ where} \sum_{\gamma \in \{\uparrow, \downarrow\}} p_\gamma = 1
\end{equation}
and the probabilities $p_\gamma$ are given by
\begin{equation}
    \label{weights}
    p_\gamma =  \expect{\psi_{GS}}{P_\gamma}{\psi_{GS}}.
\end{equation}
The state $\kett{\psi_\gamma(0)} = \kett{\psi_\gamma}$ then undergoes a unitary evolution with the Hamiltonian~\eqref{eq:hamiltonian}:
\begin{equation}
    \kett{\psi_\gamma(t)} = e^{-itH} \kett{\psi_\gamma(0)} = U(t)\kett{\psi_\gamma(0)}.
\end{equation}
where we have introduced the time evolution operator $U(t) = \exp(-itH)$.
To compute the quantum state evolution, we expand $U(t)$ in terms of the Chebyshev polynomials~\cite{hernandez_chebyshev,yuan2010modeling}:
\begin{equation}
  U(t) = \text{exp} \left(-i \tau \cal{G}\right) = \sum_{k=0}^{\infty}  \alpha_k (-i)^k J_k(\tau) T_k(\cal{G}),   \label{eq:Chebyshev}
\end{equation}
where $T_k$ are the Chebyshev polynomials of order $k$, $J_k(\tau)$ are the Bessel functions, $\tau = (E_\text{max} - E_\text{min}) t/2$, $\alpha_k = 1$ for $k = 0$ and $\alpha_k = 2$ for $k \geq 1 $, and $\cal G$ is a re-scaled Hamiltonian defined as
\begin{equation*}
    {\cal G} = \frac{2H - (E_\text{max}+E_\text{min})}{2(E_\text{max}-E_\text{min})}
\end{equation*}
with $E_\text{min}$ and $E_\text{max}$ being the minimal and maximal eigenvalues of $H$, respectively.
The spectrum of $\cal G$ then lies in the range $[-1, \,1]$ which ensures the convergence of the series~\eqref{eq:Chebyshev}.

High accuracy of this method is guaranteed by the super-exponential decay of the Bessel functions, and truncating the series at 40 Chebyshev polyniomals lead to numerically exact results for the time intervals considered in this paper.

\section{Robustness}

Let us now discuss the results of numerical simulations.
We fix the magnetic field $B = 0.5$ such that the ground state lies deep in the skyrmionic phase.
A projective measurement is then performed on the center site at time $t = 0$, and the two possible outcomes are $|\psi_\uparrow (0)\rangle$ and $|\psi_\downarrow (0)\rangle$.
The Hamiltonian remains unchanged, and the states undergo a time evolution for around $300 \; D^{-1}$.

\begin{figure*}
    \centering
    \includegraphics[width=0.8\textwidth]{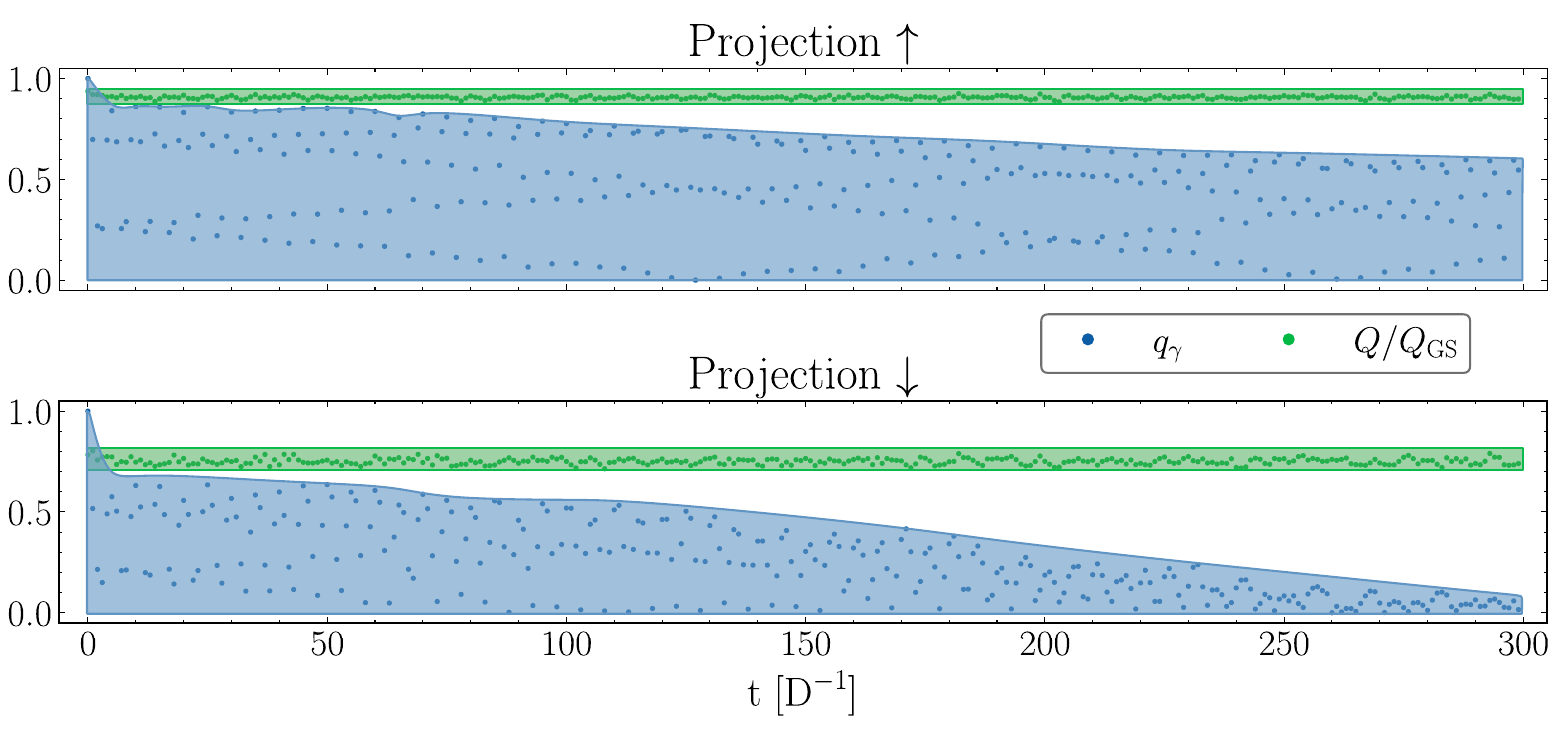}
    \caption{Time evolution of the chirality $Q$ (dark green dots) and overlap $q_\gamma$ (blue dots) after one projective measurement. The top panel shows the projection onto $|\!\uparrow\rangle$ and the bottom panel---onto $|\!\downarrow\rangle$. Lines are guides to the eye.}
    \label{fig:overlap_1meas}
\end{figure*}

In Fig.~\ref{fig:overlap_1meas}, we show the time dependence of two quantities.
First, for both states, we compute the time-dependent expectation value of the chirality~\eqref{eq:chirality} normalized by the ground state chirality $Q_\mathrm{GS} = \langle \psi_{GS} | \hat{Q} | \psi_{GS} \rangle$.
Second, we analyze how much the quantum states change in time by computing the overlap $q_\gamma(t)$ between the state at time $t$ with the state right after the projection:
\begin{equation}
    q_\gamma(t) = |\langle \psi_\gamma(0) \,|\, \psi_\gamma(t) \rangle| \,,\text{ where }\gamma \in \{\uparrow, \downarrow\}\,.
    \label{eq:overlap-single}
\end{equation}

Here, $\kett{\psi_\gamma(0)}$ should not be confused with $\kett{\psi_{GS}}$: the former denotes the state right after the projection, whereas the latter is the state before the projection.

For both $\kett{\psi_\uparrow(t)}$ and $\kett{\psi_\downarrow(t)}$, the chirality, which is shown with dark green dots, decreases slightly right after the measurement and then flattens, oscillating within a narrow range of values.
The small fluctuations around the average value hint towards spin decoherence waves propagating through the system after the measurement~\cite{first_decoherence_wave, hamieh_decoherence, hylke_decoherence, chen_spinwaves}.

The overlap $q_\gamma (t)$ exhibits highly nontrivial dynamics for both projection outcomes.
Once perturbed by the local projective measurement, the system can eventually strongly deviate from the original skyrmion ground state, while approximately retaining its spin texture as encoded in $Q$.
This is the first indication that quantum skyrmions indeed demonstrate stability in a certain sense.
Although the external perturbation destroys the original state, throughout the evolution, the system remains within a manifold of states that, on the level of operator expectation values, can be regarded as the quantum skyrmion phase.
In the following sections, we will examine this phenomenon in more detail.

\section{Multiple measurements and the quantum Zeno effect}

In the previous section, we have shown that the quantum skyrmion phase is robust with respect to a single local measurement.
A natural question to ask is what happens to the quantum system when multiple measurements are performed sequentially: will the measurements eventually destroy the quantum skyrmion, or will they keep the system pinned to the manifold of states that can be regarded as the skyrmion phase?
Here, we answer this question by systematically analyzing the effect of up to three consecutive measurements.

We start with the same ground state at $B=0.5$ and perform up to three projective measurements of the $\sigma_z$-component of the central spin.
The measurements are separated by a time interval $\delta t$ during which the system undergoes the unitary evolution with Hamiltonian~\eqref{eq:hamiltonian}.
We analyze how the resulting dynamics depends on the value of $\delta t$.
As before, we start by discussing the scalar chirality $Q$.
\begin{figure}[t!]
	\centering
	\includegraphics[width=0.9\columnwidth]{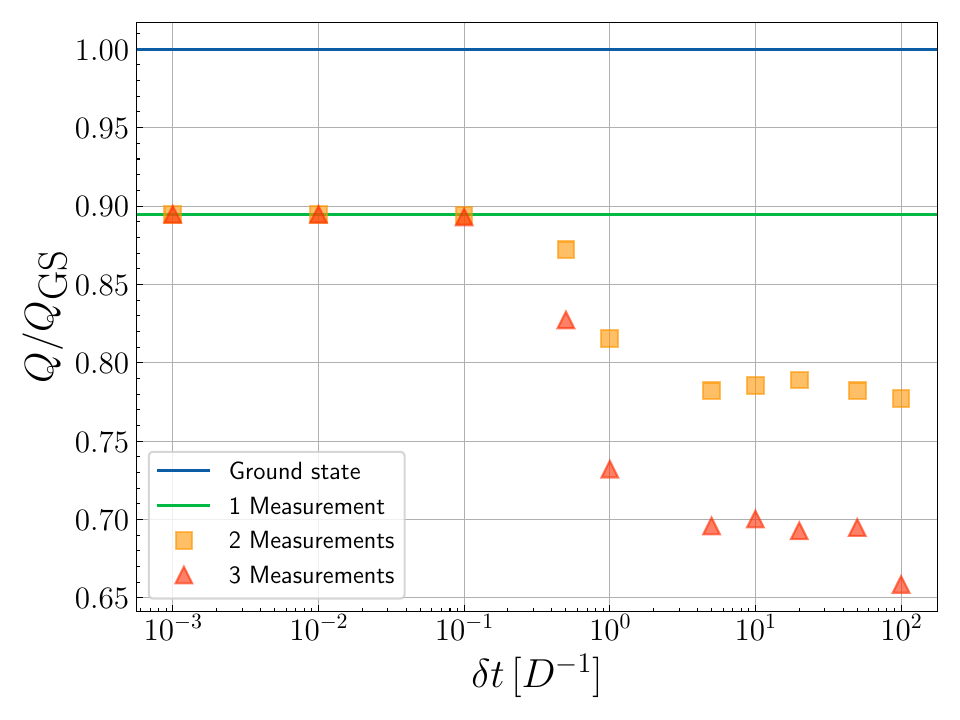}
	\caption{Averaged normalized chirality for one and repetitive measurements as a function of time interval $\delta t$ between measurements. Quantum Zeno effect stabilizes the quantum skyrmion, when projective measurements are performed for time intervals below $\delta t \leq 0.1 D^{-1}$.}
	\label{fig:chirality_multimeas}
\end{figure}

Here, we compute $Q$ immediately after the last measurement and average over all the possible outcomes of the measurements:

\begin{equation}
    \begin{aligned}
    Q &= \Tr \left[\hat{Q}\rho \right], \\
    \rho &= \sum\limits_{\gamma_i \in \{\uparrow,\downarrow\}^{\otimes i}} p_{\gamma_i} | \psi_{\gamma_i}\rangle \langle \psi_{\gamma_i}| \,, \label{eq:vonNeumann_sequence}
    \end{aligned}
\end{equation}
where $\rho$ is the von Neumann density matrix of rank $2^i$, similar to Eq.~\eqref{eq:density_operator}, and $|\psi_{\gamma_i}\rangle$ is the state immediately after the sequence of measurements $\gamma_i$.
The probabilities $p_{\gamma_i}$ are given by the products of the individual outcome probabilities.
For instance, $p_{\uparrow\downarrow\downarrow} = p_{1,\uparrow}\cdot p_{2,\downarrow}\cdot p_{3,\uparrow}$.

In Fig.~\ref{fig:chirality_multimeas}, we show $Q/Q_{GS}$ as a function of the time interval $\delta t$.
For $\delta t> 0.1 \,D^{-1}$, each subsequent measurement decreases the value of $Q$ by around $10\%$ of its original ground state value, which implies that the quantum skyrmion is not robust upon multiple perturbations, if they occur rarely enough.
However, for shorter time intervals between the measurements, $\delta t\leq 0.1 \,D^{-1}$, the subsequent measurements after the first one do not affect the value of $Q$.

\begin{figure}[t!]
	\centering
	\includegraphics[width=.48\textwidth]{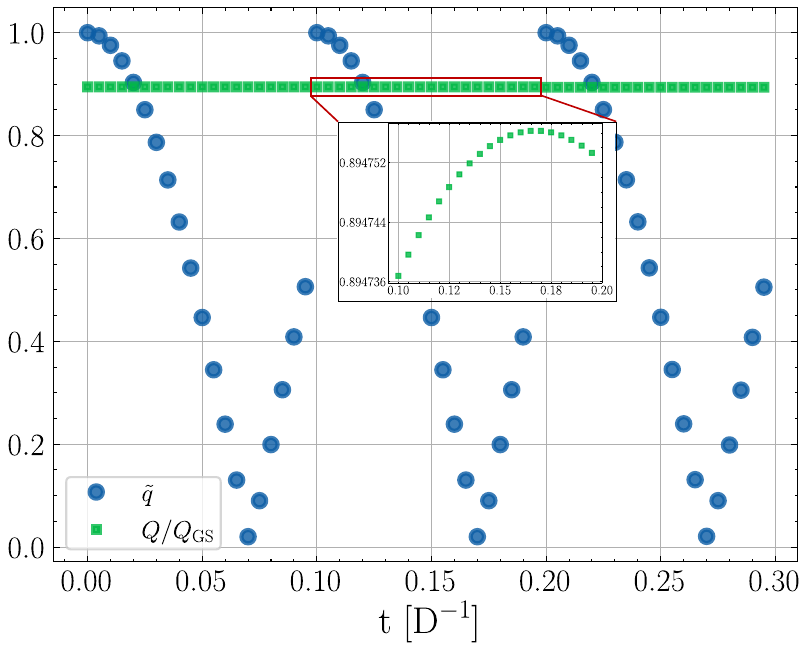}
	\caption{Time evolution of the scalar chirality $Q/Q_{GS}$ (green squares) and the overlap $\tilde{q}$ (blue circles) for three consecutive measurements.
            The measurements take place at $t=0$, $t=0.1\,D^{-1}$, and $t=0.2\,D^{-1}$. The inset zooms in on the oscillations of the chirality.
        }
	\label{fig:overlap_multimeas}
\end{figure}

Naively, one may suppose that the chirality expectation value does not really change in the regime of short inter-measurement intervals because the quantum state does not have time to significantly evolve between the measurements.
To test whether this is the case or whether the roots of the robustness of $Q$ are less trivial, consider the evolution of the wave function.
For that, we define a weighted sum of overlaps as
\begin{equation}
    \tilde{q}(t) = \left\{
        \begin{aligned}
            \sum\limits_{\gamma_1\in \{\uparrow,\downarrow\}^{\otimes 1}} &p_{\gamma_1}\cdot q_{\gamma_1}(t), & t \in [0, \delta t] \,, \\
            \sum\limits_{\gamma_2\in \{\uparrow,\downarrow\}^{\otimes 2}} &p_{\gamma_2}\cdot q_{\gamma_2}(t-\delta t), & t \in [\delta t, 2 \delta t] \,, \\
            \sum\limits_{\gamma_3\in \{\uparrow,\downarrow\}^{\otimes 3}} &p_{\gamma_3}\cdot q_{\gamma_3}(t-2\delta t), & t \in [2\delta t, 3 \delta t] \,, \\
            &\dots
        \end{aligned}
    \right.
    \label{overlap_avg}
\end{equation}
Here, $q_{\gamma_i}(t)$ is defined just as in Eq.~\eqref{eq:overlap-single}, but with multiple consecutive projections.
For $\delta t = 0.1\, D^{-1}$, dynamics of this overlap is shown in Fig.~\ref{fig:overlap_multimeas} alongside the evolution of the scalar chirality.
Note, that after each measurement, the overlap is reset to $1$ because we compute it with respect not to the original ground state, but to the state right after the corresponding projective measurement.
It can be seen that the weighted overlap quickly deviates from $1$, and the time-evolved states can essentially become orthogonal to the states of the system right after the measurement.
After the first measurement, the relative change of $Q$ is negligibly small, around $10^{-5}$.
In other words, the chirality is affected by neither measurements nor the interim unitary evolution.
We suggest that this can be interpreted as a certain type of the Quantum Zeno Effect (QZE).
In its simplest form, QZE means that a quantum system with a discrete spectrum that is monitored continuously (or, in practical terms, frequently enough) with von Neumann projective measurements cannot undergo a transition to another state even when its initial state is not stable~\cite{sudarshan_zeno}.
Here, it is not quite the case since, after each measurement, the system rapidly evolves away from the initial state, and the subsequent measurements do not bring it back.
However, a broader class of QZE-like phenomena allow the system to non-trivially evolve in such a way that its state remains within a Zeno subspace defined by the measurement~\cite{facchi_zeno, aftab_zeno}, and the studied case appears to fall into this category.
In what follows, we will examine the process of Zeno stabilization in more detail and argue that QZE can be used to amplify the robustness of quantum skyrmions.

\section{Further manifestations of the Quantum Zeno Effect and skyrmion's robustness}

\subsection{Low-energy excitation spectrum}

For the QZE scenario to be relevant, the quantum system's dynamics should involve states belonging to a discrete part of the energy spectrum.
Although, formally speaking, the spectrum of a 19-spin system is discrete, some parts of it might transform into continuous bands upon increasing the system size to the thermodynamic limit.
To ensure that the measurement-driven dynamics of the quantum skyrmion occurs outside of such quasi-continuous parts of the spectrum, we study the spectrum of the system in more detail.

\begin{figure}[t!]
	\centering
	\includegraphics[width=0.8\columnwidth]{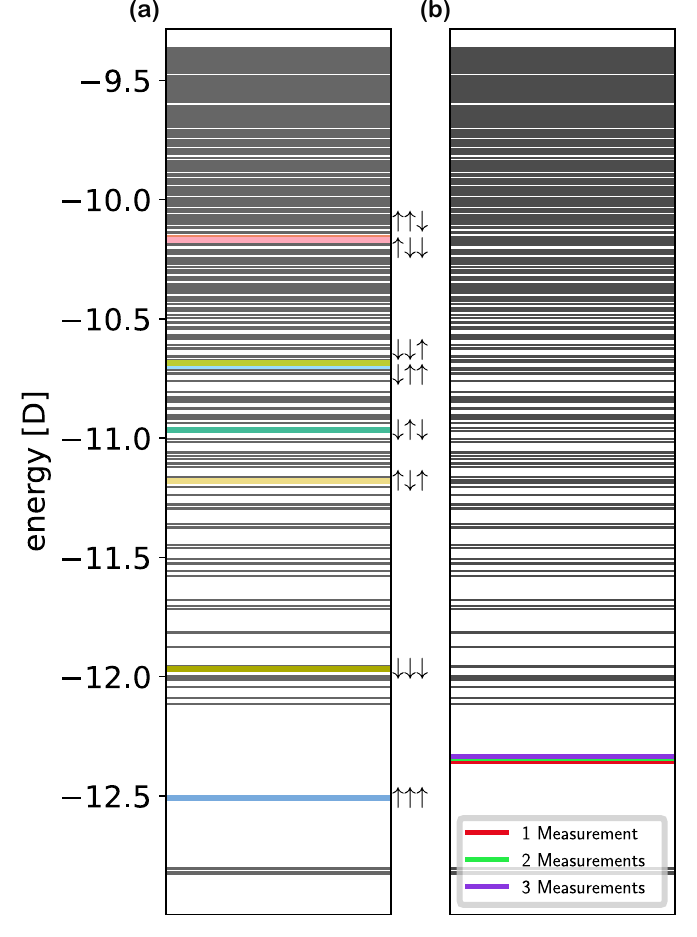}
	\caption{The lowest part of the energy spectrum of the system.
            Energy levels are indicated with thin black lines.
            (a) Energies of different outcomes of projective measurements.
            (b) Energy of the system (i.e., weighted average the possible outcomes) for one, two, and three consecutive measurements.}
	\label{fig:energy_meas}
\end{figure}

We use the lattice-symmetries~\cite{westerhout2021lattice} package to reconstruct the lowest-lying $\sim\!\! 1500$ eigenstates.
The energy spectrum is shown in Fig.~\ref{fig:energy_meas}.
The spectrum has a large gap between a few nearly degenerate low-lying states and the quasicontinuum of higher excitations.
Projective measurements drive the system from the ground state into a higher-energy state, and the energy after the measurements can fall into either the gap or the quasicontinuum, depending on the specific measurement outcomes, as shown in Fig.~\ref{fig:energy_meas}a.
However, the energy averaged by the von Neumann density matrix~\eqref{eq:vonNeumann_sequence} remains within the gap, Fig.~\ref{fig:energy_meas}b.
Hence, the quantum skyrmion can still be discussed within the quantum Zeno effect paradigm.

\subsection{Longitudinal spin structural factor}

Although the scalar chirality $Q$ is the main characteristic feature of the quantum skyrmion phase, it cannot be probed experimentally due to the complexity of measuring three-point correlation functions.
An alternative signature of a quantum skyrmion is the spin structural factor~\cite{haller_skyrm}
\begin{equation}
    X^\parallel_\mathbf{q} = \average{{S}^z_\mathbf{q}{S}^z_\mathbf{-q}} \,,
    \label{eq:XSq}
\end{equation}
where $\mathbf{q}$ is scattering wave vector. $X^\parallel_\mathbf{q}$ can be measured in neutron scattering experiments~\cite{Lovesey}, and in this section we analyze its robustness upon local projective measurements.

\begin{figure*}
	\centering
	\includegraphics[width=\textwidth]{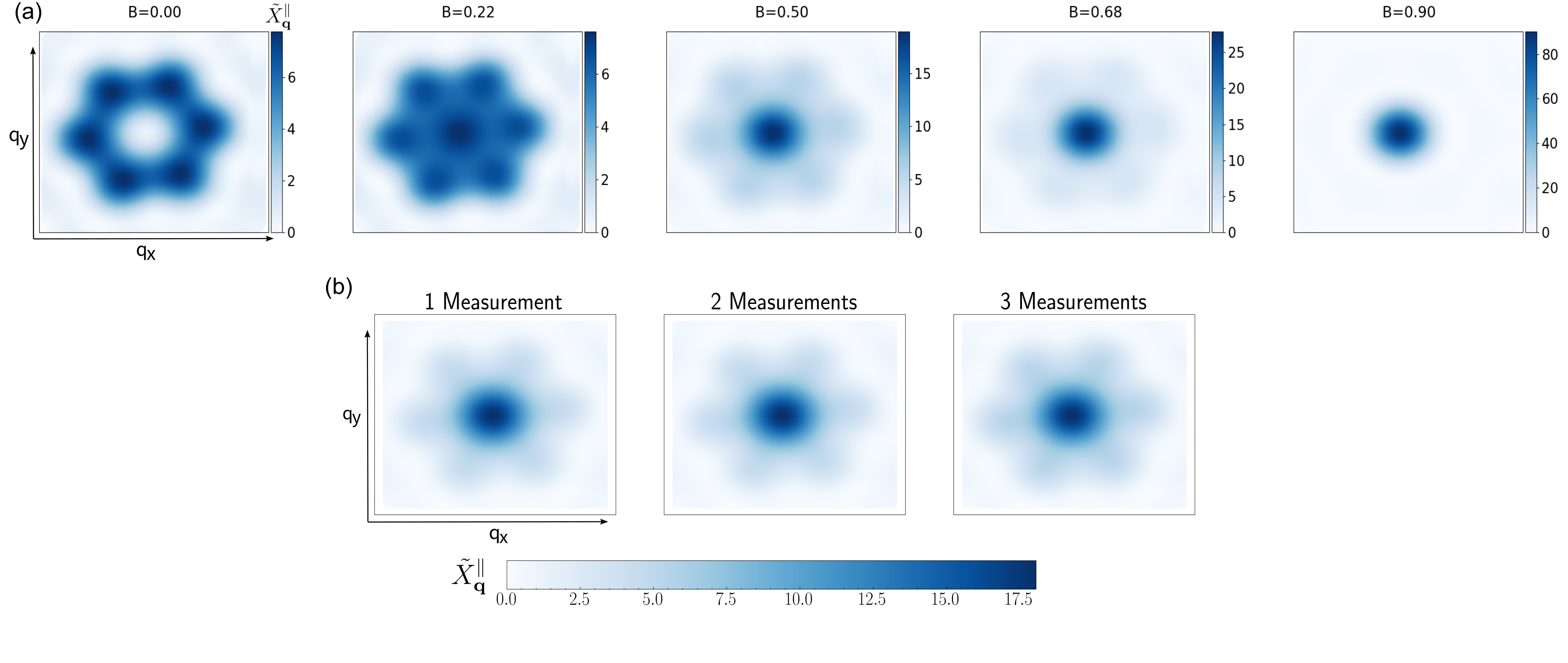}
	\caption{The longitudinal spin structural factor of the system.
            (a) $X^\parallel_\mathbf{q}$ in the ground state for various values of the magnetic field $B$.
            (b) $\tilde{X}^\parallel_\mathbf{q}$ immediately after one, two, and three measurements. The value of the magnetic field was $B=0.5$, and the time interval between the measurements was $\delta t = 0.1\, D^{-1}$.
        }
	\label{fig:sslf}
\end{figure*}

In Fig.~\ref{fig:sslf}a, we show the intensity profile of the spin structural factor~\eqref{eq:XSq} calculated in the ground state of the Hamiltonian~\eqref{eq:hamiltonian} for several different values of the magnetic field.
Although in Ref.~\cite{sotnikov_skyrm}, it was argued that $X^\parallel_\mathbf{q}$ is a sub-optimal probe of quantum skyrmion phase due to the lack of clear difference between the skyrmionic and helical states,
here we see that the skyrmion has a recognizable pattern of Bragg peaks characterized by a strong zero-momentum peak and a halo of weaker peaks.
If the pattern remains unchanged after a sequence of projective measurements, it would speak in favor of the operational robustness of quantum skyrmions.

The spin structural factor averaged over the measurement outcomes is
\begin{equation}
    \tilde{X}^\parallel_\mathbf{q} = \sum\limits_{\gamma_i} p_{\gamma_i} \langle \psi_{\gamma_i}| {S}^z_\mathbf{q}{S}^z_\mathbf{-q}|\psi_{\gamma_i}\rangle
\end{equation}
In Fig.~\ref{fig:sslf}b, we show this quantity after one, two, and three projective measurements.
The profile of the structural factor remains nearly unchanged, with only a slight shift of intensity of the halo.
Hence, we can safely claim that, on the level of observables (the scalar chirality and the spin structural factor), the quantum skyrmion phase is robust upon external local perturbations despite the absence of topological protection.

\section{Conclusions}

We studied the robustness of a quantum skyrmion state with respect to local projective measurements using a 19-site triangular lattice as a model system.
First, we analyzed the effect of the local projective measurements on the scalar chirality, a three-spin correlation function that can be considered a quantum analog of the skyrmionic topological charge~\cite{sotnikov_skyrm,mazur_jpsj}.
The scalar chirality was not a topological charge, and its robustness with respect to perturbations was not guaranteed.
Nevertheless, we demonstrated that a local projective measurement had a relatively weak effect on the scalar chirality and reduced it by only around 10\%.

Second, we analyzed the effect of up to three consecutive projective measurements and found that it depended on the time interval between the measurements.
If the interval was small enough ($\lesssim 0.1\;D^{-1}$), an analogue of the quantum Zeno effect arose, and the skyrmion phase was stabilized by the measurements.
We discussed the physical origin of this behavior by analyzing the energy transfer to the system during the projective measurements.

Finally, in the quantum Zeno regime, we also demonstrated the robustness of the spin structural factor as a quantity easier accessible experimentally than the scalar chirality.

A physical realization of the quantum skyrmion phase would open up a way to experimentally study subtle effects of quantum measurements,
and the operational robustness upon local projective measurements suggests that the study of the quantum skyrmion phase is a promising research direction with potential applications in reading and writing of information.

\section*{Acknowledgements}
This work was supported by the European Research Council (ERC) under the European Union's Horizon 2020 research and innovation program, grant agreement 854843-FASTCORR. The data that support the findings of this study are available from the corresponding author upon reasonable request.

\bibliography{biblio_quantumskyrm}

\end{document}